\documentclass[%
 amsmath,amssymb,
 aps,
]{revtex4-1}

\usepackage{graphicx}
\usepackage{dcolumn}
\usepackage{bm}

\begin{document}

\title{$W-$hairs of black holes in three dimensional spacetime}
\author{Jingbo Wang}
\email{ shuijing@mail.bnu.edu.cn}
\affiliation{Institute for Gravitation and Astrophysics, College of Physics and Electronic Engineering, Xinyang Normal University, Xinyang, 464000, P. R. China}
 \date{\today}
\begin{abstract}
In the previous paper (arXiv:1804.09438) we found that the near horizon symmetry algebra of black holes is a subalgebra of the $W_{1+\infty}$ symmetry algebra of quantum Hall fluid in three dimensional spacetime. In this paper, we give a slightly different representation of the former algebra from the latter one. Similar to the horizon fluff proposal, based on the $W_{1+\infty}$ algebra, we count the number of the microstates of the BTZ black holes and obtain the Bekenstein-Hawking entropy.
\end{abstract}
\bibliographystyle{unsrt}
\maketitle
\section{Introduction}
Information paradox\cite{info1} is still a challenging problem in theoretical physics. It says that the evaporation of black hole will break the unitary. Up to now, there are many proposals for its resolution. For some reviews, see Ref.\cite{info2,info3}. In 2016, Hawking, Perry and Strominger\cite{softhair1} suggest to use ``soft hair" to solve this paradox. They proposed that the black hole microstates could be related to the soft hairs, that is, the zero energy excitations on the horizon. Since then, there are a lot of works along this direction, see Ref.\cite{softhair2,softhair3} and references therein.

A related concept, named ``horizon fluff", was present in Ref.\cite{fluff1,fluff2,fluff3}. Based on the a new near horizon boundary condition, a new near horizon symmetry, which is infinite copies of the Heisenberg algebra was obtained. The horizon fluff forms a finite subset of the related ``soft Heisenberg hair". Use this algebra one can generate descendants of physical states which are interpreted as black hole microstates in three dimensional spacetime. The number of those microstates counts for the entropy of the black holes.

In previous paper \cite{wangti3}, we establish that the near horizon symmetry algebra of three-dimensional black holes \cite{nhs1} is a subalgebra of the $W_{1+\infty}$ algebra of quantum Hall fluid. The $W_{1+\infty}$ algebra is the quantum version of area-preserving diffeomorphism algebra, which is a dynamical symmetry of quantum Hall liquid. Based on this result, we give a classification of black holes. They are specified by two integers $(n_1,n_2)$ which are functions of the black hole mass and angular momentum $(M,J)$. In this paper, we gives a different embedding of near horizon symmetry algebra into the $W_{1+\infty}$ algebra, which is more similar to the horizon fluff proposal. Based on this algebra we give the ``W-hairs" \cite{whair1} of the BTZ black holes. Actually the $W_{1+\infty}$ algebra was used to retain the information in two-dimensional stringy black holes \cite{whair2,whair3,whair4,whair5}. This algebra also appears in the spectrum of Hawking radiations \cite{hawkradia1,hawkradia21,hawkradia2,hawkradia3,hawkradia4,hawkradia5}.

The paper is organized as follows. In section II, the representation of the $W_{1+\infty}$ for black holes is outlined. The embedding of the near horizon symmetry algebra into the $W_{1+\infty}$ algebra is obtained. In section III, following the horizon fluff proposal, the W-hairs of the BTZ black hole is proposed, and it can explain the Bekenstein-Hawking entropy. Section IV is the conclusion.
\section{Near horizon symmetry algebra from $W_{1+\infty}$ symmetry algebra}
The generators $V_n^i$ of the $W_{1+\infty}$ are characterized by a mode index $n \in Z$ and a conformal spin $h=i+1$, and satisfy the algebra \cite{wshen},
\begin{equation}\label{2}
  [V_n^i, V_m^j]=(j n-i m)V_{n+m}^{i+j-1}+q(i,j,n,m)V_{n+m}^{i+j-3}+\cdots+c(n)\delta^{ij} \delta_{m+n,0},
\end{equation}
where $q(i,j,n,m)$ are pertinent polynomials and $c(n)$ represents the relativistic quantum anomaly. The dots stand for a series of terms involving the operators $V_{n+m}^{i+j-1-2k}$.

The generators $V_n^0$ and $V_n^1$ form a subalgebra of the $W_{1+\infty}$ algebra:
\begin{equation}\label{3}\begin{split}
  [V^0_n, V^0_m]&=n c \delta_{n+m,0},\\
  [V^1_n, V^0_m]&=-m V^0_{n+m},\\
  [V^1_n, V^1_m]&=(n-m)V^1_{n+m}+\frac{c}{12} n(n^2-1)\delta_{n+m,0},\\
\end{split}\end{equation}
with central charge $c=1$. It contains an abelian Kac-Moody algebra and $c=1$ Virasoro algebra.

All unitary, irreducible, highest-weight representations have been found by Kac and Radul \cite{kr1,kr2}. This result was applied to incompressible quantum Hall fluid by Cappelli et al \cite{ctz1,ctz2}. These representations exist only for positive integer central charge $c=m=1,2,\cdots$. If $c=1$, they are equivalent to those of the Abelian subalgebra $\widehat{U(1)}$ of $W_{1+\infty}$, corresponding to the edge excitations of a single Abelian Chern-Simons theory. For $c=m=2,3,\cdots$, there are two kinds of representation, generic and degenerate, depending on the weight. The generic representations are equivalent to the corresponding representations of the multi-component Abelian algebra $\widehat{U(1)}^m$ which corresponds to the edge excitations of a multiple Abelian Chern-Simons theory. On the other hand, the degenerate representations are contained in the $\widehat{U(1)}^m$ representations.

Any unitary, irreducible representation contains a bottom state--the highest-weight state, and an infinite tow(descendants) above it. The highest-weight state $|\Omega>$ is defined by the conditions
\begin{equation}\label{4}
  V_n^i |\Omega>=0, \quad \forall n>0, i\geq 0.
\end{equation}
Applying polynomials of $V_n^i (n<0)$ on the $|\Omega>$ will give other excitations.

It was claim that the black holes can be considered as a quantum spin Hall state in three dimensional spacetime \cite{wangti1,wangti2}. A quantum spin Hall state can be realized as a bilayer integer quantum Hall system with opposite $T-$ symmetry. So the algebra for a quantum spin Hall state is $W_{1+\infty} \otimes \bar{W}_{1+\infty}$, which have opposite chirality. For integer quantum Hall fluid, the representation is $c=1$, thus $\widehat{U(1)}$ algebra. So for black holes, the corresponding algebra is $W=\widehat{U(1)}\otimes \widehat{\bar{U}(1)}$ which have opposite chirality. This result can also be obtained from the Chern-Simons theory \cite{whcft1}.

Now we considered the representation of this algebra $W=\widehat{U(1)}\otimes \widehat{\bar{U}(1)}$ \cite{ctz3}. Firstly consider the chiral part $\widehat{U(1)}$. The generators $\alpha_n^+$ satisfy
\begin{equation}\label{5}
[\alpha^+_n, \alpha^+_m]=n \delta_{n+m,0}.
\end{equation}
All $V_n^i$ can be written as polynomials of the current modes $\alpha^+_n$.

All unitary, irreducible representations can be built on the top of highest-weight state $|r_1>, r_1 \in R$, which satisfies
\begin{equation}\label{6}
 \alpha^+_n |r_1>=0 \quad (n>0), \quad  \alpha^+_0 |r_1>=r_1 |r_1>.
\end{equation}
A general descendant can be written as
\begin{equation}\label{7}
  |\{n_1,n_2,\cdots,n_s\}>=\alpha^+_{-n_1} \alpha^+_{-n_2} \cdots \alpha^+_{-n_s} |r_1>,  \quad n_1\geq n_2 \geq \cdots \geq n_s>0.
\end{equation}
Notice that the operator $\alpha^+_0$ commutate with all other generators, which means that the eigenvalues of $\alpha^+_0$ are the same for all descendants in a given representation.

The Virasoro generator $L_n^+$ can be defined through Sugawara construction
\begin{equation}\label{8}
  L_n^+=\frac{1}{2}\sum_{l \in Z}:\alpha^+_{n-l} \alpha^+_l:,
\end{equation}
where $::$ is normal ordering. Acting on the highest-weight state gives
\begin{equation}\label{9}
 L^+_n |r_1>=0 \quad (n>0), \quad  L^+_0 |r_1>=\frac{r_1^2}{2}|r_1>.
\end{equation}
They satisfy the Virasoro algebra in (\ref{3}) with $c=1$.

Similarly we consider the anti-chiral part. The generators are $\bar{\alpha}_n^+$ satisfying
\begin{equation}\label{10}
[\bar{\alpha}^+_n, \bar{\alpha}^+_m]=-n \delta_{n+m,0}.
\end{equation}
The highest-weight state $|r_2>, r_2 \in R$ is defined by
\begin{equation}\label{11}
 \bar{\alpha}^+_n |r_2>=0 \quad (n<0), \quad  \bar{\alpha}^+_0 |r_2>=r_2 |r_2>.
\end{equation}
The Virasoro generators $\bar{L}_n^+$ can also be defined through Sugawara construction (\ref{8}), but sadly they do not satisfy the Virasoro algebra (\ref{3}). It is possible to define the new operators
\begin{equation}\label{12}
\alpha^-_n \equiv \bar{\alpha}^+_{-n}, \quad L^-_n \equiv -\bar{L}^+_{-n},
\end{equation}
which indeed satisfy the standard algebras (\ref{3}),(\ref{8}) and conditions (\ref{6}),(\ref{9}).

At last we get two copies of $\widehat{U(1)}$ algebra,
\begin{equation}\label{13}
[\alpha^\pm_n, \alpha^\pm_m]=n \delta_{n+m,0},
\end{equation}
which is the same as the algebra in Ref.\cite{fluff1} except the insignificant factor $1/2$.

With those algebra one can construct the near horizon symmetry algebra. Define
\begin{equation}\label{14}
  T_n=\alpha^+_n+\alpha^-_{-n},\quad Y_n=L^+_n-L^-_{-n}.
\end{equation}
It is easy to show that those operators satisfy the algebra \cite{nhs1}
\begin{equation}\label{1}\begin{split}
  [T_m, T_n]&=0,\\
  [Y_m, T_n]&=-nT_{m+n},\\
  [Y_m, Y_n]&=(m-n)Y_{m+n}.
\end{split}\end{equation}
The $T_n$ generates a supertranslation and $Y_n$ generates a superrotation.
\section{W-hairs of BTZ black holes}
In this section we discuss the representations of the algebra (\ref{13}). According to the rule of conformal field theory, those representations should be closed under the ``fusion algebra". For $\widehat{U(1)}$ it is just the adding of $r$. For  $W=\widehat{U(1)}\otimes \widehat{\bar{U}(1)}$, the hight-weight states can be written as $|r_1,r_2>,r_1,r_2 \in R$. The operators (\ref{14}) acting on this state give
\begin{equation}\label{15}
  T_0 |r_1,r_2>=(r_1+r_2)|r_1,r_2>,\quad Y_0 |r_1,r_2>=\frac{r^2_1-r^2_2}{2}|r_1,r_2>,
\end{equation}
A general descendant can be written as
\begin{equation}\label{16}
  |\{n_i^\pm\}>=\prod_{n_i^\pm} (\alpha^+_{-n_i^+} \alpha^-_{-n_i^-})|r_1,r_2>,  \quad n^\pm_1\geq n^\pm_2 \geq \cdots \geq n^\pm_s>0.
\end{equation}
The operators on those states give
\begin{equation}\label{17}
  T_0  |\{n_i^\pm\}>=(r_1+r_2) |\{n_i^\pm\}>,\quad Y_0  |\{n_i^\pm\}>=(\frac{r^2_1-r^2_2}{2}+\sum n_i^+-\sum n_i^-) |\{n_i^\pm\}>.
\end{equation}
The key problem is to choose which representations correspond to the BTZ black holes. Following the horizon fluff proposal, we made the following assumption: the BTZ black holes correspond to the descendants of the absolute vacuum state $|(r_1=0,r_2=0)>$. Thus the BTZ black hole states can be written as \cite{fluff1}
\begin{equation}\label{17a}
|B\{n_i^\pm\}>=N\{n_i^\pm\} \prod_{n_i^\pm}(\alpha^+_{-n_i^+} \alpha^-_{-n_i^-})|0,0>,  \quad n^\pm_1\geq n^\pm_2 \geq \cdots \geq n^\pm_s>0,
\end{equation}
which $N\{n_i^\pm\}$ is the normalization factor.

It is useful to compare with the quantum Hall fluid. For QHE, the $T_0$ represent the electric charge and $Y_0$ the angular momentum of the quasi-particles. For black holes, the meaning of $T_0$ is unclear, but the $Y_0$ indeed represents the angular momentum. Let's define another operator $H=L^+_0+L^-_{0}$, which is the Hamiltonian. Then we identify the BTZ black hole with parameters $(M,J)$ with the descendant $|B\{n_i^\pm\}>$ which satisfies
\begin{equation}\label{18}
  <B'|Y_0|B>=c J \delta_{B',B},\quad <B'|H|B>=c M l \delta_{B',B},
\end{equation}
where $c=3l/2G$ is the red-shift factor near the horizon \cite{fluff3}. Submitting (\ref{17a}) into (\ref{18}) gives
\begin{equation}\label{19}
  \sum n_i^+-\sum n_i^-=c J,\quad \sum n_i^++\sum n_i^-=c M l.
\end{equation}
The solution is simple,
\begin{equation}\label{20}
 \sum n_i^+=c \frac{M l+J}{2}, \quad  \sum n_i^-=c \frac{M l-J}{2}.
\end{equation}
Different $\{n_i^\pm\}$ corresponds to different microstate of the BTZ black hole. The total number of the microstates for BTZ black hole with parameters $(M,J)$ is given by the famous Hardy-Ramanujan formula,
\begin{equation}\label{20a}
  p(N)\simeq \frac{1}{4N \sqrt{3}}\exp(2 \pi \sqrt{N/6}).
\end{equation}
The entropy of BTZ black hole is given by the logarithm of the number of microstates $|B\{n_i^\pm\}>$,
\begin{equation}\label{21}
  S=\ln p(c \frac{M l+J}{2})+\ln p(c \frac{M l-J}{2})+\cdots=\frac{2 \pi r_+}{4G}+\cdots,
\end{equation}
which is just the Bekenstein-Hawking entropy with some corrections of order $\ln S$.

The next question is what the other highest-weigh states $|r_1,r_2>,r_1,r_2 \in R$ mean. Let us turn back to quantum Hall fluid again. In quantum Hall fluid, there are two kinds of excitations: the neutral excitations, and charged excitations which corresponds to quasi-holes and quasi-particles in the bulk of the fluid. For fully-filled Landau level which is integer quantum Hall effect, the highest-weigh state is the vacuum state $|0>$ and the descendants above it. For fractional quantum Hall effect the other highest-weigh states $|Q>$ appear which has fractional charges and statistics. In black hole side, correspondingly, the pure black holes, associated with absolute vacuum state $|0>$, and black holes interacting with matters may corresponds to other highest-weigh states.
\section{Conclusion}
In this paper, we consider the $W_{1+\infty}$ symmetry algebra in three dimensional spacetime. For black holes, the corresponding algebra is $W=\widehat{U(1)}\otimes \widehat{\bar{U}(1)}$ which have opposite chirality. The explicit form is given in (\ref{13}). From this algebra one can easily get the near horizon symmetry algebra (\ref{1}).

The infinite set of W charges provide an infinite set of discrete gauge hairs (W-hairs) \cite{whair1}, which were used to maintain the quantum coherence for two-dimensional stringy black hole. In this paper, we associate those W-hairs with the microstates of black holes, following the sprit of horizon fluff. The BTZ black holes can be considered as the descendants of the absolute vacuum state, i.e. (\ref{17a}). For BTZ black hole with parameter $(M,J)$ we can count the number of those microstates to gives the Bekenstein-Hawking entropy. The essential difference with the horizon fluff proposal is that we use the $W_{1+\infty}$ algebra instead of the Heisenberg algebra, even through for black holes case their representations are very similar.

The near horizon symmetry algebra is related to the fluid symmetry algebra in Ref.\cite{penna1}. In this paper, we give an explicit fluid, the quantum Hall fluid. The microscopic structure of this fluid is well understood. The relation between those two infinite-dimensional algebras also give another evidence to support our claim that ``black hole can be considered as kind of topological insulator". This claim relate the black hole physics with the condensed matter physics. It is also the starting point to relate the gravity with some non-trivial condensed matter systems \cite{hu1,volo1,lib1,vaid1}.
\acknowledgments
 This work is supported by the NSFC (Grant No.11647064) and Nanhu Scholars Program for Young Scholars of XYNU.

\bibliography{bms2}
\end{document}